# Decoherence during time evolution by Schroedinger equation

Shunichi Muto

Department of applied physics, Graduate School of Engineering, Hokkaido University, Sapporo, Hokkaido, 060-8628, Japan

Abstract

A thought experiment is discussed to clarify the concept of decoherence. Superposition of states consisting of ground state of a single hydrogen atom and its excited state after a huge amount of time is discussed to show that the decoherence of the state, more precisely the loss of coherence, is literally the loss of ability of interference. The loss of superposition of states is just apparent or even superficial. Applications of this example in its relation to many-world interpretation of observation of a quantum state, decoherence in usual cases, and weak collapse of wavepackets are discussed.

Introduction

Concept of decoherence, or loss of coherence is a fundamental subject of quantum mechanics. However, there still seems to be a misunderstanding coming from complexities such as interactions with environments, scattering, and observation. Therefore, we discuss here one of the simplest cases of decoherence which lacks any explicit mechanism of decoherence. The system we treat here is a half-excited hydrogen atom in a huge universe. We show that, even in this simple system, decoherence is realized after a long time and this is done without destroying the superposition of states.

1. Thought experiment and results

Suppose we have a single hydrogen atom in a large universe (Fig. 1). Let's assume that the atom is initially in the ground state with energy $-E_1$ where $E_1$ is Rydberg energy. We excite the atom with coherent laser pulse with energy $h\nu$. If the pulse is a $\pi/2$ pulse to excite the ground state to a state with energy $h\nu - E_1$. and if the pulse duration, $\Delta t$ is long enough to define the excited state, in other words, $\hbar/\Delta t \ll h\nu, E_1$, then the result is an equal superposition of 2 states ; the ground state and the excited continuum state with kinetic energy $E_2 = h\nu - E_1$. The excited state is not stable and it travels isotropically with a speed $v_g$ of about $\hbar k/m$, where $m$ is the electron mass and $k$ is the $k$-vector of the electron given by the kinetic energy: $\hbar^2 k^2/2m = E_2$.

Suppose we wait for a long time $T$ after the excitation such that $v_g T$ is on the order of the size of universe $L$. This means that $T$ is order of $L/v_g$ or $Lm/\hbar k$. After $T$, the coherence of two states are completely lost. Electron in the ground state stays in the hydrogen atom which is at rest and electron in the excited state is far away in the universe. There is no way those states can interfere with each other*.

Note that the coherence is lost without any decoherence mechanism. The whole process is completely describable by simple time-dependent Schroedinger eq. Therefore, the superposition of states is completely reserved. The density matrix now is given by

$$p = 1/2|1><1| + 1/2|2><2| \qquad (1).$$

Here $|1>$ and $|2>$ indicate the ground and excited states. The reason that the cross terms $|1><2|$ and $|2><1|$ do not appear is because there is no operator which can connect those two state, in other words, for any operator A,

$$<1|A|2> = <2|A|1> = 0 \qquad (2),$$

and any physical quantity is calculated by trace, $Tr \rho A = A_{11}\rho_{11} + A_{22} \rho_{22}$. Here $\rho_{11}$ and $\rho_{22}$ are 1/2 for this case of $\pi/2$ pulse.

There is one thing to be noted here. The excited state in this case is not exactly a simple state. Exactly it consists of combination of states with energy around $E_2$. Especially the lifetime $\delta t$ of the state around the hydrogen atom gives the range of energy contributing to this "state". This is determined by $l/v_g = lm/\hbar k$, where $l$ is Bohr radius or the range of the electron localized in the ground state. The width of energy $\delta E$ due to this lifetime is $\delta E = \hbar/\delta t = \hbar^2 k/lm$. Comparing this width with the average energy, $\delta E/E = 2/\lambda k = 1/\pi \times \lambda_e/l$. Here, $\lambda_e$ is the wavelength of the travelling electron. If $\lambda_e$ is much smaller than Bohr radius, the energy witdth is negligible compared with the total energy $E_2$ or $h\nu$. Then we can treat the excited state as a state with approximate energy $E_2$. This ambiguity of the excited state could be improved by introducing elaborate structures to allow virtual bound state which replaces the excited state. This can be done by replacing the hydrogen atom by an impurity or quantum dot in semiconductor heterostructures. However, essence is still the same and we are not going to discuss those now.

We assumed here that the energy ambiguity coming from the finite duration of π-pulse is negligible. Numerically, if $E_2$ is 100 eV, $k$ is $4.9 \times 10^{10}$ /m and $\delta t$ is $1.7 \times 10^{-17}$ s or 17 atoseconds. $v_g$ is $5 \times 10^6$ m/s and $T$ is as large as $10^{12}$ years for $L$ of $137 \times 10^8$ light year. If we compromise $L$ to be 1 light year, then $T$ is rather short (about 60 years).

2. Discussions

The result illusturated by eq. (1) shows that the superposition of states is not lost. What is lost is the coherence, or, more precisely, the possibility of interference of two states. Therefore, decoherence is the loss of literal coherence and not the superposition. The loss of superposition is just apparent. When we observe the hydrogen atom, in other words, the presence and absence of electron around the proton, we find which of the world of |1> and |2> we are in.

What is learned from the above "gedanken" experiment for real world? Usually on earth, we do not have this kind of size infinity. Instead we have infinity in the degrees of freedom or the dimension of Hilbert space. Since we have this kind of infinity, we can easily lose coherence of objects due to interactions with emvironments and scatterings. These interactions, scattering, and infinite degrees of freedom obscure the essence of physical process. However, as in the simple case we disucussed above, the superposition is researved. Especially in observation of superposed states, interaction with measurement apparatus or "detector" leads to superpositon of states which is infinitely appart from each other in the huge detector Hilbert space. This seems to support the original interpretaion of observation problem of quantum mechanics by Everett [1] "whatever the wavefunction describes is actually realized". This resulted in famous Many-Worlds Interpretation of Quantum Mechanics by DeWitt [2]. Therefore, the "collapse" or "reduction" of wavepacket is not needed to understand the quantum mechanics.

Finally some remarks on weak collapse of wavepacket. If a quantum measurement is done to a superposed state and nobody see the apparatus which shows the result of measurement. This is sometimes called a weak collapse of wavepacket. The density matrix to describe this situation is analogous to the one given by eq. (1). However, the resultant state is just like the result of our "gedanken" experiment and the process is completely described by Schroedinger eq. Therefore, there is neither "weak collapse" nor "collapse" of wavefunction. One last remark of the applicability of our discussions. For whole of the above discussions we neglected the effect of gravity and everything is within the framework of linear Schroedinger eq. Therefore, if quantum gravity is properly formulated someday, it is possible, though not likely, that the above argument collapses.

In summary, superposition of states consisting of ground state of a single hydrogen atom and it's excited state after a huge amount of time is discussed. It is shown that the decoherence of the state, or, more precisely, the loss of coherence, is the loss of possibility of interference and not the loss of superposition. People may say that there is nothing new in this thought experiment. However, the author hopes that this may be helpful in clarifying the essence of decoherence, to especially those who get lost in huge Hilbert space.

Acknowledgments

The author would like to thank all students participated in my freshmen class of

"quantum world" in 2011 and 2012. The discussions with them encouraged him very much to finish this manuscript. Also, he would like to thank Dr. S. Tanda for pointing out the negligence of gravity in the arguments.

## Figure caption

Figure 1.  A half-excited hydrogen atom

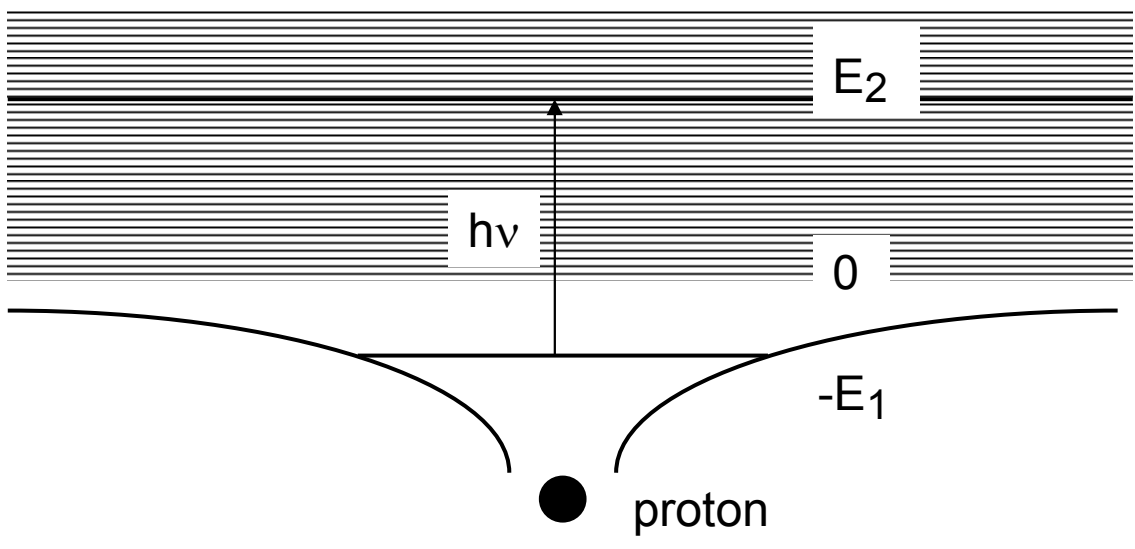

Fig. 1    A half excited hydrogen atom